\documentclass[mypaper]{myaa}  
\usepackage{graphicx}
\usepackage{balance}
\usepackage{txfonts}
\usepackage{natbib}
\usepackage[figuresright]{rotating}
\usepackage{aalongtable}
\def\kms{$\mathrm {km s}^{-1}$}            
\idline{0}{1}

\begin{document}
   \title{The exotic chemical composition of the Sagittarius dwarf Spheroidal galaxy}

   \author{L. Sbordone	
          \inst{1,2,4}
	\and
          P. Bonifacio
	  \inst{1,2,3}
	\and
	  R. Buonanno
	  \inst{5,8}
	\and
	  G. Marconi
	  \inst{6}
	\and
	  L. Monaco
	  \inst{6}
	  \fnmsep
	\and
	  S. Zaggia
	\inst{7}
          }

   \offprints{L. Sbordone}

   \institute{
         CIFIST Marie Curie Excellence Team
         \and
         Observatoire de Paris, GEPI , 5, place Jules Janssen, 92195 Meudon, France\\
         \email{luca.sbordone@obspm.fr}
         \and
         INAF -- Osservatorio Astronomico di Trieste -- via G. B. Tiepolo 11, 34131 Trieste, Italy
         \and
         INAF -- Osservatorio Astronomico di Roma -- via di Frascati 33, 00040 Monte Porzio Catone, Rome, Italy
	 \and
	 Universit\`a di Roma ``Tor Vergata'', via della Ricerca Scientifica 1, 00133 Rome, Italy
	 \and
	 ESO -- European Southern Observatory, Alonso de Cordova 3107, Santiago, Chile	 
         \and
         INAF -- Osservatorio Astronomico di Padova, Vicolo dell'Osservatorio 5, 35122 Padua, Italy 
         \and
         ASI Science Data Center, via Galileo Galilei, 00044 Frascati, Italy
             }

   \date{}

  \abstract
   {The \object{Sagittarius dwarf Spheroidal Galaxy} is the nearest neighbor of the Milky Way. 
Moving along a short period quasi-polar orbit within the Halo, 
it is being destroyed by the tidal interaction with our Galaxy, losing its stellar content along a huge stellar stream.}
   {We study the detailed chemical composition of 12 giant stars in the Sagittarius dwarf Spheroidal main body, together with 5 more in the associated globular cluster \object{Terzan 7} by means of high resolution VLT-UVES spectra.}
   {Abundances are derived for up to 21 elements from O to Nd, by fitting lines EW or line profiles against ATLAS 9 model atmospheres and SYNTHE spectral syntheses calculated ad-hoc. Temperatures are derived from (V-I)$_0$ or (B-V)$_0$ colors, gravities from \ion{Fe}{i} - \ion{Fe}{ii} ionization equilibrium.}
   {The metallicity of the observed stars is between [Fe/H]=-0.9 and 0. We detected a highly peculiar ``chemical signature'', with undersolar $\alpha$ elements, Na, Al, Sc, V, Co, Ni, Cu and Zn among others, and overabundant La, Ce and Nd. Many of these abundance ratios (in particular light-odd elements and 
iron peak ones) are strongly at odds with what is observed within the Milky Way, they thus may be 
a very useful tool to recognize populations originated within the Sagittarius dwarf. This can be clearly seen in the case of the globular \object{Palomar 12}, which is believed to have been stripped from Sagittarius: the cluster shows precisely the same chemical ``oddities'', thus finally confirming its extragalactic origin.}
   {}


\titlerunning{Chemical composition of Sgr dSph}

   \maketitle

\section{Introduction}

Dwarf Spheroidal Galaxies (dSph) have become in recent years a highly popular subject of investigation. The interest towards such objects has been largely driven by the key role they are supposed to play in the buildup process of larger galaxies such as the Milky Way. The available detailed abundance ratios for stars in Local Group (LG) dSph are nevertheless showing that {\em present day} dSph are undesirable candidates as hierarchical merging building blocks \citep[see][and references therein]{vladilo,venn04}. The distinctive abundance ratios (most notably, the low [$\alpha$/Fe] ratio) observed in the LG dSph hint for star formation histories remarkably different from the one characteristic of the MW, thus making it difficult, for {\em evolved} dSphs to have played a significant role in building up our Galaxy. Such findings are, after all, not surprising: undisturbed, low mass dSph are a different environment from the MW, where low star formation rates and highly efficient galactic winds have likely played a major
role \citep[][]{lanfranchi03,lanfranchi04,lanfranchi05}. Dwarf Spheroidals are nevertheless not ruled out as MW ``building blocks'', but the main merging phase should have taken place at a very early stage, allowing the subsequent evolution to differentiate the surviving dSph from the larger, merged structures.

Nevertheless, at least one major merging episode is currently taking place in the MW, at expenses of the Sagittarius dSph \citep[Sgr dSph,][]{ibata94,ibata95}. The nearest known dSph \citep[26.3 Kpc,][]{monaco04}, \object{Sgr dSph} is being tidally destroyed while moving along its quasi-polar, short period (less than 1 GYr) orbit around the MW \citep{ibata97,helmi99}, and its stars are dispersing along a huge stream in the Halo {\bf \citep{majewski03,belokurov06}}.

In previous works \citep[][henceforth paper I, II and III respectively]{bonifacio00,bonifacio04,sbordone05} we presented chemical abundance for total 12 stars in the \object{Sgr dSph} main body and in the associated globular cluster \object{Terzan 7}. Paper I reported abundances for 20 elements from O to Eu in two \object{Sgr dSph} main body stars, paper II added Fe and $\alpha$ elements for 10 more stars, while paper III analyzed Fe, $\alpha$ elements and Ni in 5 giants in \object{Terzan 7}. In two more papers, we have also obtained iron and $\alpha$ elements abundances for 15 brighter RGB stars of Sgr \citep[][]{monaco05}, and Sulphur abundances for three stars in \object{Ter 7} \citep{caffau05}. The present paper unifies, extends and revisits  the results presented in Paper I, II and III: 

\begin{itemize}
\item the temperature scale has been recalibrated for the 12 \object{Sgr dSph} main body stars, by using photometries and reddening from \citet{monaco02}. This has been made to homogenize the temperature scale with the one used in \citet{monaco05}. See section \ref{observations} for details;
\item the array of abundances has been extended to up to 21 species including O, Na, Mg, Al, Si, Ca, Sc, Ti, V, Cr, Mn, Fe, Co, Ni, Cu, Zn, Y, Ba, La, Ce, and Nd;
\item with respect to the previous papers, updated atomic data are used for Mg, Ca, La, Ce, and Nd.
\end{itemize}

\section{Observations, data reduction and analysis}
\label{observations}

As above stated, the present work uses the same data employed in Paper I and II for the \object{Sgr dSph} main body, and in Paper III for \object{Terzan 7} stars. The reader is thus referred there for the details of the observations. Table \ref{pos_and_phot} lists photometry and derived atmospheric parameters for the 12 \object{Sgr dSph} main body and the 5 \object{Ter 7} giants, a sample of the spectra for the 12 \object{Sgr dSph} stars is shown in fig. 1 of Paper II, and in fig. 1 of Paper III for the 5 \object{Ter 7} giants.
The temperature scale was recalibrated in \object{Sgr dSph} main body due to the adoption of \citet{monaco02} photometry and \citet{layden00} reddening, while in Paper I and II we used \citet{marconi98} photometry and reddening estimates. This change came from the need of using a temperature scale homogeneous across our present work, and applicable to future studies also. Moreover \citet{monaco02} photometry covers a much larger field and it was used as a basis for FLAMES candidates selection \citep[see][]{zaggia04}.  

Effective temperatures were derived from dereddened (V-I) colors by means of the \citet{alonso99,alonso01} calibration for giant stars. The change in the photometry and reddening correction (E(V-I)=0.22 in \citealt{marconi98} were substituted by E(V-I)=0.18 from \citealt{layden00}\footnote{{\bf Reddening estimate from \citet{schlegel98} is very near to the \citet{layden00} one, with tipical value of  E(V-I)=0.19 for our Sgr dSph main body stars}}. led to significantly {\em lower} derived effective temperatures, with a mean decrease of about 250 K. The reason for such a change comes partly from the different reddening estimates (0.04 difference in V-I color), but is mainly due to an offset between the two photometries (0.074 mean). {\bf As a further test, we checked the temperatures derived from 2MASS J-K colors for stars in this work and in \citet{monaco05}. Nevertheless, the Sgr dSph stars presented here are generally too faint for 2MASS to provide reliable colors for them, and the derived temperatures were scattered over more than 1000 K. For the \citet{monaco05} stars, 2MASS J-K colors led to temperatures in average 75 K {\em hotter}  than the ones based on \citet{monaco02} V-I colors, while on Ter 7 there was no systematic discrepancy between the two temperature scales.} We will show further on (see \ref{errors}) how the changes in the physical parameters of the atmospheres do not alter significantly the ``scientific output'' of this research.

Similarly to what we did in paper I through III, one-dimensional, LTE atmosphere models were computed for the observed stars by means of our GNU-Linux ported version of the ATLAS code. ATLAS 9 (using Opacity Distribution Functions, henceforth ODF) models were used for all the stars except for star $\# \object{1665}$ in \object{Terzan 7} which required an ATLAS 12 (opacity sampling) model due to its very low temperature and gravity \citep[see ][for details]{sbordone05}. For the ODF based models, no $\alpha$-enhancement was assumed (since [$\alpha$/Fe] appears to be solar or sub-solar); ``new'' type ODF \citep{castelli03} were used.  Abundances were computed from measured line equivalent widths (EW) by means of WIDTH, model gravity was set by imposing \ion{Fe}{i} -- \ion{Fe}{II} ionization equilibrium, and microturbulence by requesting abundances of \ion{Fe}{i} lines to be independent of the line EW. Abundances for lines affected by strong hyperfine splitting, or for which EW measurement was problematic (e. g. the 630 nm OI line) were derived by spectral synthesis, using SYNTHE \citep[for ATLAS, WIDTH and SYNTHE see][]{kurucz93,kurucz05,sbordone04,sbordone05b}. 

To produce the same \ion{Fe}{ii} line strength with a lower T$_{eff}$, lower gravity is needed. Our gravity estimates decreased by about 0.3 -- 0.4 dex with respect to the ones in paper II, leading to values of $\log g \sim 2.1$. This caused the compatibility between isochrone and ionization gravities to worsen somewhat in comparison with Paper I and II: superimposing \citet{girardi02} isochrones of compatible age and metallicity leads now to derive typical gravities of the order of $\log g \sim 2.4$, for a mean discrepancy of about 0.35 dex.

\begin{table}[t]
\begin{center}
\caption{Photometry and atmospheric parameters for the studied stars. Employed colors are (V-I){\boldmath $_0$} for \object{Sgr dSph} main body and (B-V){\boldmath $_0$} for \object{Ter 7}. Coordinates can be found in Paper II (\object{Sgr dSph} main body) and Paper III (\object{Ter 7}). \label{pos_and_phot}}
\begin{tabular}{llllll}
\hline
{\bf Star} & {\bf V} & {\bf (V-I)}{\boldmath $_0$} &
{\bf T}{\boldmath $_{eff}$} & {\boldmath $\log{g}$} & {\boldmath $\xi$} \\ 
 & mag & mag & K & cgs & \kms \\ 
\hline
\\
Sgr \object{432} & 17.700 & 1.013 & 4713 & 2.2   & 1.20  \\
Sgr \object{628} & 18.228 & 1.040 & 4656 & 2.1   & 1.75  \\
Sgr \object{635} & 18.186 & 1.074 & 4588 & 2.1   & 1.50  \\
Sgr \object{656} & 18.217 & 1.008 & 4723 & 2.1   & 1.50  \\
Sgr \object{709} & 18.260 & 1.034 & 4669 & 2.2   & 1.20  \\
Sgr \object{716} & 18.282 & 1.012 & 4715 & 2.1   & 1.60  \\
Sgr \object{717} & 18.282 & 1.007 & 4726 & 2.0   & 1.10  \\
Sgr \object{772} & 18.392 & 1.070 & 4596 & 1.9   & 1.60  \\
Sgr \object{867} & 18.465 & 1.031 & 4675 & 1.7   & 1.95  \\
Sgr \object{879} & 18.516 & 1.073 & 4590 & 1.9   & 1.30  \\
Sgr \object{894} & 18.507 & 1.067 & 4602 & 2.1   & 1.50  \\
Sgr \object{927} & 18.580 & 1.079 & 4578 & 2.1   & 1.30  \\
\\
         &       &{\bf (B-V)}{\boldmath $_0$} & & & \\
Ter7 \object{1272} & 16.62 & 1.15  & 4421 & 1.2   & 1.45  \\
Ter7 \object{1282} & 16.08 & 1.30  & 4203 & 1.3   & 1.60  \\
Ter7 \object{1515} & 16.76 & 1.12  & 4468 & 2.0   & 1.45  \\
Ter7 \object{1665} & 15.04 & 1.50  & 3945 & 0.8   & 1.60  \\
Ter7 \object{1708} & 16.08 & 1.28  & 4231 & 1.2   & 1.70  \\
\\
\hline
\end{tabular}
\end{center}
\end{table}

Complete line lists, with employed $\log gf$ values, measured EW and derived abundances are available in the online version. An excerpt is available in table \ref{line_list}. 
Paper I and II employed the same set of lines and atomic data (with the exception of \ion{O}{i} 630 nm line) for the elements which were common among the two. Paper III used the same set of lines of Paper I and II for the Dichroic I spectra and (obviously) a different one for the Dichroic II spectra. Nevertheless, in Paper III $\log gf$ were updated for Mg and Ca lines. Globally, updates are as follows:
\begin{itemize}
\item{\bf \ion{O}{i} } we used the \citet{storey00} $\log gf$ already from Paper II on. \ion{O}{i} 630 nm line is measured by spectral synthesis. The line is heavily blended with a weak \ion{Ni}{i} feature for which we used the \citet{johansson03} laboratory  $\log gf$;
\item{\bf \ion{Mg}{i} } from Paper III on, where applicable we use the \citet{gratton03} $\log gf$;
\item{\bf \ion{Si}{i} } solar \citet{edvardsson93} $\log gf$ values are used where applicable;
\item{\bf \ion{Ca}{i} } since Paper III we use, where present, \citet{smith81} furnace $\log gf$; 586.7 nm line has \citet{gratton03} $\log gf$;
\item{\bf \ion{Zn}{i} } we used the $\log gf$ values of \citet{biemont80} which provide a good agreement between the photospheric and the meteoritic Zn abundance;
\item{\bf \ion{La}{ii} } recent $\log gf$ values from \citet{lawler01} are used here when available;
\item{\bf \ion{Ce}{ii} } $\log gf$ values are taken from \citet{hill95};
\item{\bf \ion{Nd}{ii} } new $\log gf$ values are taken from \citet{denhartog03};
\end{itemize}
All the other atomic data are the ones provided in the line lists included with our ATLAS suite port \footnote{Available at {\tt http://wwwuser.oat.ts.astro.it/atmos/}}, which come from R. L. Kurucz website \footnote{{\tt http://kurucz.harvard.edu/}}. Hyperfine splitting was used to synthesize lines for \ion{Mn}{i}, \ion{Co}{i},\ion{Cu}{i}. Solar isotopic ratios were assumed. 

Many of the employed Na and Al lines are known to be significantly affected by NLTE \citep{Baumueller97, Baumueller98, Gratton99}. NLTE corrections are generally a function of metallicity, atmospheric parameters and line strength, and unfortunately no correction computations have been produced for Al lines for giant stars such the one we are dealing with. As a consequence, we publish LTE abundances for Al. For Na lines, correction are computed for giant stars by \citet{Gratton99}. Their grid extends down to $\log(g)$=1.5 and T$_{eff}$=4000 K, which fits all the \object{Sgr dSph} main body stars and one of the \object{Ter 7} stars. For these stars we interpolated the correction values and present here the corrected abundances. The other four \object{Ter 7} stars go beyond the grid in gravity or temperature, or both. In these cases, extrapolated values can be used, but given the complex behavior of NLTE corrections in phase space, we preferred to adopt a single value of A$_{NLTE}$-A$_{LTE}$=0.2.

\begin{table*}[t]
\begin{center}
\caption{An excerpt from the detailed line tables available in the Online version. Employed lines, $\log gf$ values, $\log gf$ sources, observed EW and derived abundances for \object{Sgr dSph} stars \# \object{432} to \#\object{716}. Codes for the source of the $\log gf$ can be found in the bibliography. For lines that have been synthesized, ``syn'' substitutes the EW. LTE abundanced are listed here for the Na lines.\label{line_list}}
{\scriptsize
\begin{tabular}{rrrlrrrrrrrrrrrrrr}
\hline\\
Ion        & $\lambda$ & $\log gf$  & source of    & EW      & A(X)       & EW       & A(X)       & EW      & A(X)       & EW      & A(X)       & EW     & A(X)       & EW     & A(X)      \\
           & (nm)      &         & $\log gf$       & (pm)    &            & (pm)     &            & (pm)    &            & (pm)    &            & (pm)   &            & (pm)   &           \\
           &           &         &              & \object{432}     &            & \object{628}      &            & \object{635}     &            & \object{656}     &            & \object{709}    &            & \object{716}    &           \\
\\
\hline
\\
\ion{O }{i} & 630.0304 & -9.717  & STZ          &   syn   &  8.03      &   syn    & 8.35       &   syn    & 8.33      &  syn    &  8.15      & syn    &  8.50      &  syn   &  8.54     \\
\ion{Na}{i} & 568.2633 & -0.700  & KP           &   --    &  --        &  10.67   & 5.79       &   7.56   & 5.33      &   9.61  &  5.76      & 11.23  &  6.06      & 11.18  &  5.94     \\
\ion{Na}{i} & 615.4227 & -1.560  & KP           &   --    &  --        &   4.38   & 5.72       &   1.95   & 5.21      &   3.36  &  5.62      &  --    &  --        &  2.47  &  5.44     \\
\ion{Na}{i} & 616.0747 & -1.260  & KP           &   --    &  --        &   5.26   & 5.56       &   3.60   & 5.26      &   4.34  &  5.49      &  6.07  &  5.74      &  5.77  &  5.69     \\
\hline\\
\end{tabular}
}
\end{center}
\end{table*}

\section{Results}
\label{results}

Absolute abundances for \object{Sgr dSph} stars, solar assumed abundances, [Fe/H], [X/Fe] and associated errors are listed in table \ref{abs_abu_sgr_1} and \ref{abs_abu_sgr_2}. The same data for \object{Terzan 7} stars are in table \ref{abs_abu_ter}. 

\subsection{The \object{Sgr dSph} main body}

As already stated in Paper I and II, the 12 \object{Sgr dSph} stars have a relatively high mean metallicity ([Fe/H] between -0.89 and 0.02): the average value is [Fe/H]=-0.36. In comparison with Paper I and II, the variaton in the atmospheric parameters led to a slight metallicity decrease.

\begin{figure}                                                                                      
\centering                                                                                          
\includegraphics[width=8.cm]{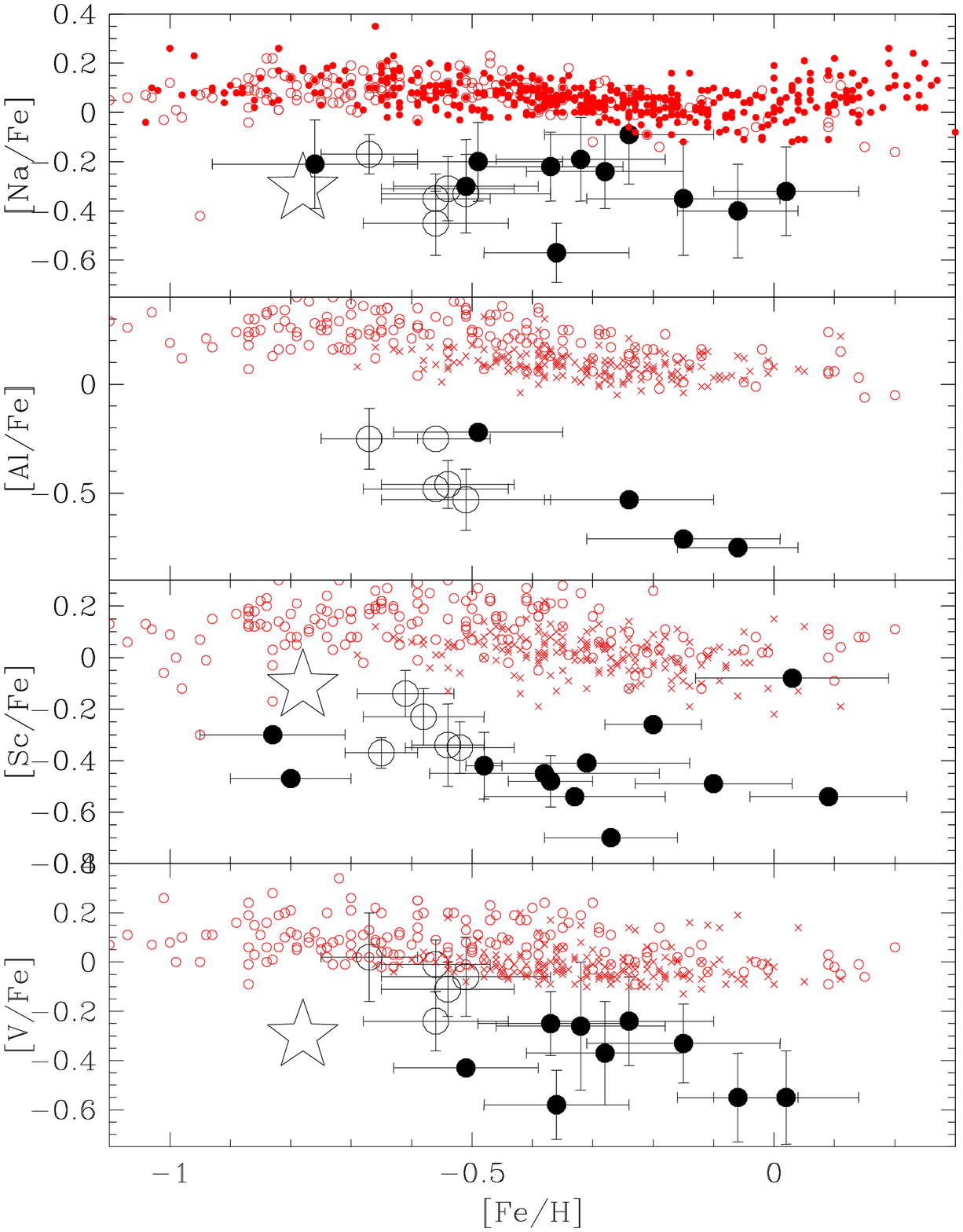}                                                       
\caption{[Na/Fe], [Al/Fe], [Sc/Fe] and [V/Fe] plotted against [Fe/H].  The symbols are as follows: large filled circles, \object{Sgr dSph} main body, large open circles \object{Ter 7}, large star, \object{Palomar 12} \citep[mean value for the stars in][]{cohen04}, small open circles, MW sample from \citet{Reddy06} small crosses, MW sample from \citet{Reddy03}, small filled circles, MW sample from \citet{venn04}, which also includes \citet{Reddy03} stars. In the Na plot, the value for \object{Pal 12} has been corrected by the same amount used for  low-gravity \object{Ter 7} stars (A$_{NLTE}$-A$_{LTE}$=0.2) \label{NaAlScV}}
\end{figure} 

   \begin{figure*}
   \centering
   \includegraphics[height=15cm,angle=270]{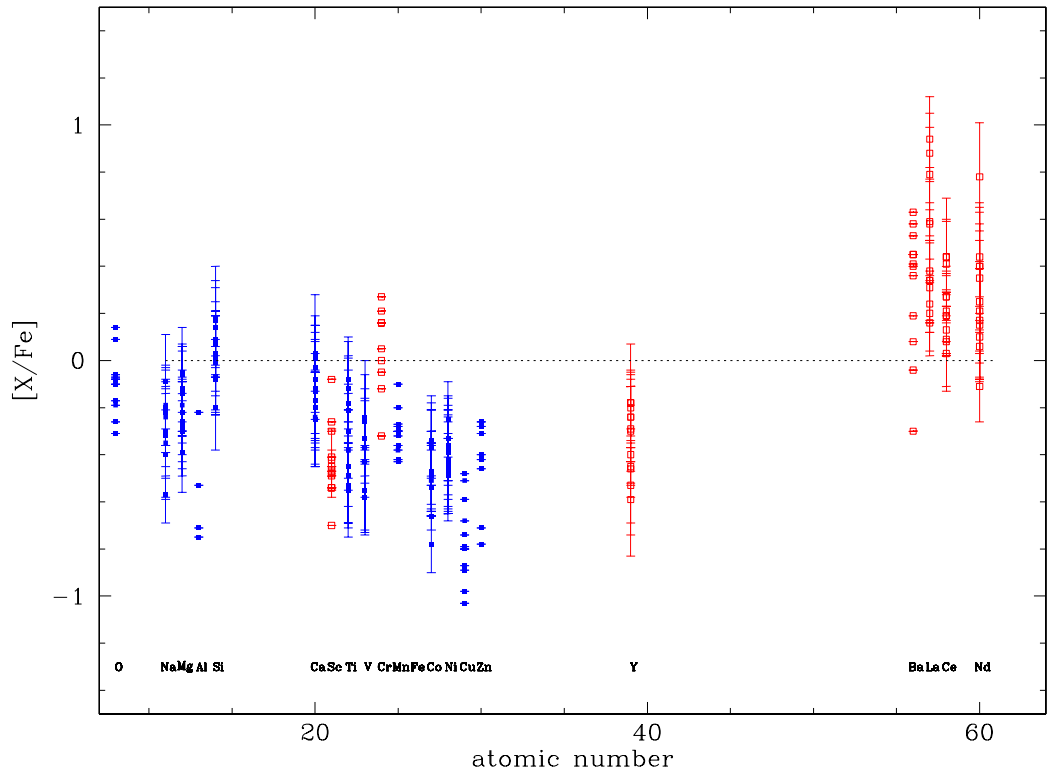}
      \caption{ The ``chemical signature'' of the 12 \object{Sgr dSph} stars: [X/Fe] ratios are plotted against atomic number. The dashed line at 0 value represents the solar abundance ratios. As in tab. \ref{abs_abu_sgr_1}, \ref{abs_abu_sgr_2} and \ref{abs_abu_ter}, ratios are against \ion{Fe}{i} for neutral species (filled squares), against \ion{Fe}{ii} for the ionized ones (Sc, Cr, Y, Ba, La, Ce, Nd, open squares). Error bars are the same listed in tab. \ref{abs_abu_sgr_1}, \ref{abs_abu_sgr_2} and \ref{abs_abu_ter}, so that species measured on a single line do not show any error bar. Important departures from solar ratios can be seen throughout all the sampled elements: see text for details. \label{signature}
              }
         
   \end{figure*}

A full picture of the ``chemical signature'' of \object{Sgr dSph} main body is presented in figure \ref{signature}. At first glance, \object{Sgr dSph} appears to bear the signs of a highly peculiar chemical evolution. We will now treat in some detail the various element groups.

\begin{figure*}                                                                                  
   \centering                                                                                       
      \includegraphics[height=15.cm,angle=270]{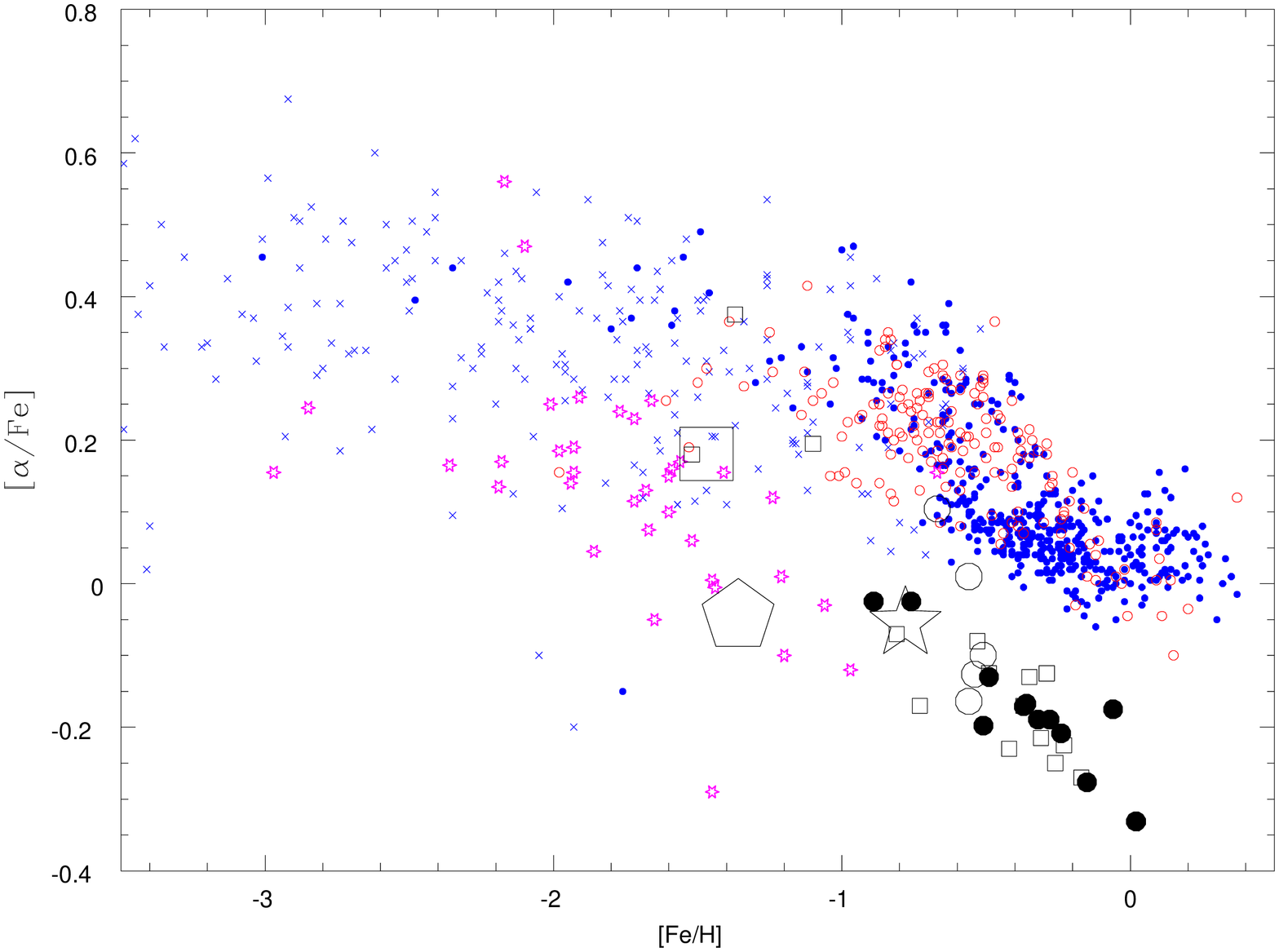}                              
      \caption{ [$\alpha$/Fe] (defined as mean of [Mg/Fe] and [Ca/Fe]) is plotted against [Fe/H] for various samples: large filled dots, \object{Sgr dSph} main body (this work); 
      large open dots, \object{Ter 7} (this work); open squares, \object{Sgr dSph} main body upper RGB stars \citep{monaco05}, small open stars, LG dSph stars \citep[Car, Dra, For, Leo I, 
      Scl, Sex, UMi,][]{shetrone01, shetrone03}. Small symbols refer to MW samples: filled circles, \citet{venn04} thin and thick disk stars; crosses, \citet{venn04} Halo
      stars; open circles, \citet{Reddy06}, mainly thick disk stars. Huge open symbols refer to mean values for globular clusters: star, Palomar
      12 \citep[4 stars,][]{cohen04}; square \object{M 54} \citep[5 stars,][]{brown99}; pentagon, \object{Ruprecht 106} \citep[2 stars,][]{brown97}. \object{Ru 106} is included due to its low
      [$\alpha$/Fe], [Ni/Fe] ratios and its high radial velocity, hinting for an extragalactic origin, although its association with \object{Sgr dSph} is unlikely \citep[see][]{pritzl05}. \label{alphasufe_tutti}
              }
   \end{figure*}

The $\alpha$ elements show the same behavior already described in Paper I and II (fig. \ref{alphasufe_tutti}). The $\alpha$ elements show solar or undersolar ratio against iron, with a significant trend with the metallicity, leading in particular Mg and Ca to reach heavy underabundances in the most metal rich stars of the sample.([Mg/Fe]=-0.39 in star $\# \object{709}$). Silicon remains instead nearer to solar values, and even shows some enhancement in some cases ([Si/Fe]=0.18 in star $\# \object{894}$). Titanium, \citep[although its inclusion among $\alpha$ elements is not totally correct from a nucleosynthetic point of view, see][]{b2fh} also shows heavy underabundance compared to solar values.

Light odd-atomic number elements Na, Al, Sc and V also show a highly significant underabundance with respect to solar values (see fig. \ref{NaAlScV}). A hint of a trend with metallicity can be seen in [Al/Fe] graph, but since in \object{Sgr dSph} main body stars a single line was used, and considering the possible presence of NLTE effects, this cannot be considered significant.

Iron-peak elements Co, Ni, Cu and Zn (fig. \ref{CoNiCuZn}) display perhaps the most intriguing anomalies, showing constantly undersolar ratios, with no clear metallicity trends. Although both Cu and Zn have been measured using a single line, these are rather strong and clear transitions whose fit is robust.

Finally, heavy n-capture elements Y, Ba, La, Ce and Nd (Y, Ba, La and Nd in fig. \ref{YBaLaNd}) also show interesting patterns. Y appears to be undersolar at low metallicity, with a slight increasing trend with metallicity which leads to solar [Y/Fe] ratios around solar [Fe/H]. Ba is oversolar with a moderate increasing trend with metallicity, and generally outside the range of [Ba/Fe] variation within the MW, but three \object{Sgr dSph} main body stars (\#\object{635}, \#\object{716} and \#\object{867}) show MW-like [Ba/Fe] ratios. Spectra inspection did not show clear anomalies in the Ba lines in these stars, but, since we are using a single \ion{Ba}{ii} feature, the hypothesis of some contamination of the line cannot be ruled out. Lantanium abundances show an above than average spread, but [La/Fe] appears to be consistently oversolar and increasing with metallicity, reaching rather extreme values at solar metallicity ([La/Fe]=0.94 at [Fe/H]=0.02 in star \#\object{709}). Finally, Nd starts from MW-like (essentially solar) ratios at low metallicity, increasing then slightly but never truly standing out with respect to the values observed within the MW.

\subsection{\object{Terzan 7}}

\begin{figure}               
\centering                   
\includegraphics[width=8.cm]{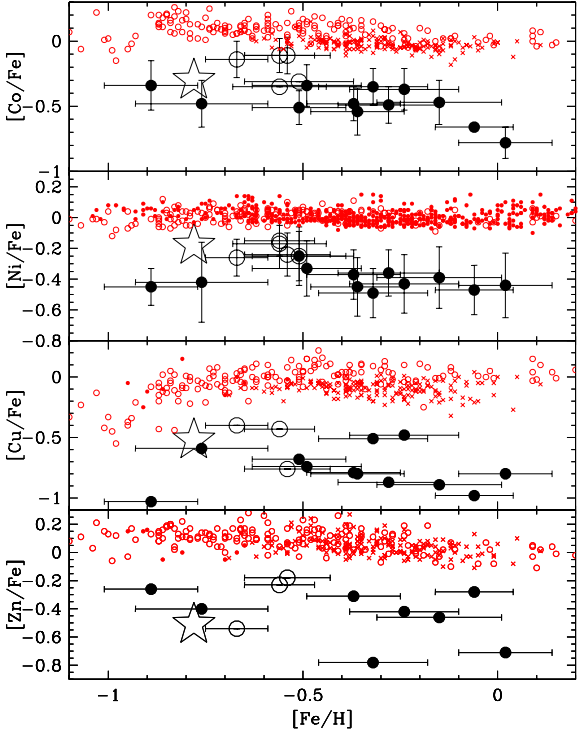}
\caption{[Co/Fe], [Ni/Fe], [Cu/Fe] and [Zn/Fe] plotted against [Fe/H].  The symbols are the same as in fig. \ref{NaAlScV}, except for small filled circles, which indicate \citet{venn04} sample for Ni, and \citet{bihain04} sample for Cu and Zn. \label{CoNiCuZn}}
\end{figure}

\begin{figure}               
\centering
\includegraphics[width=8.cm]{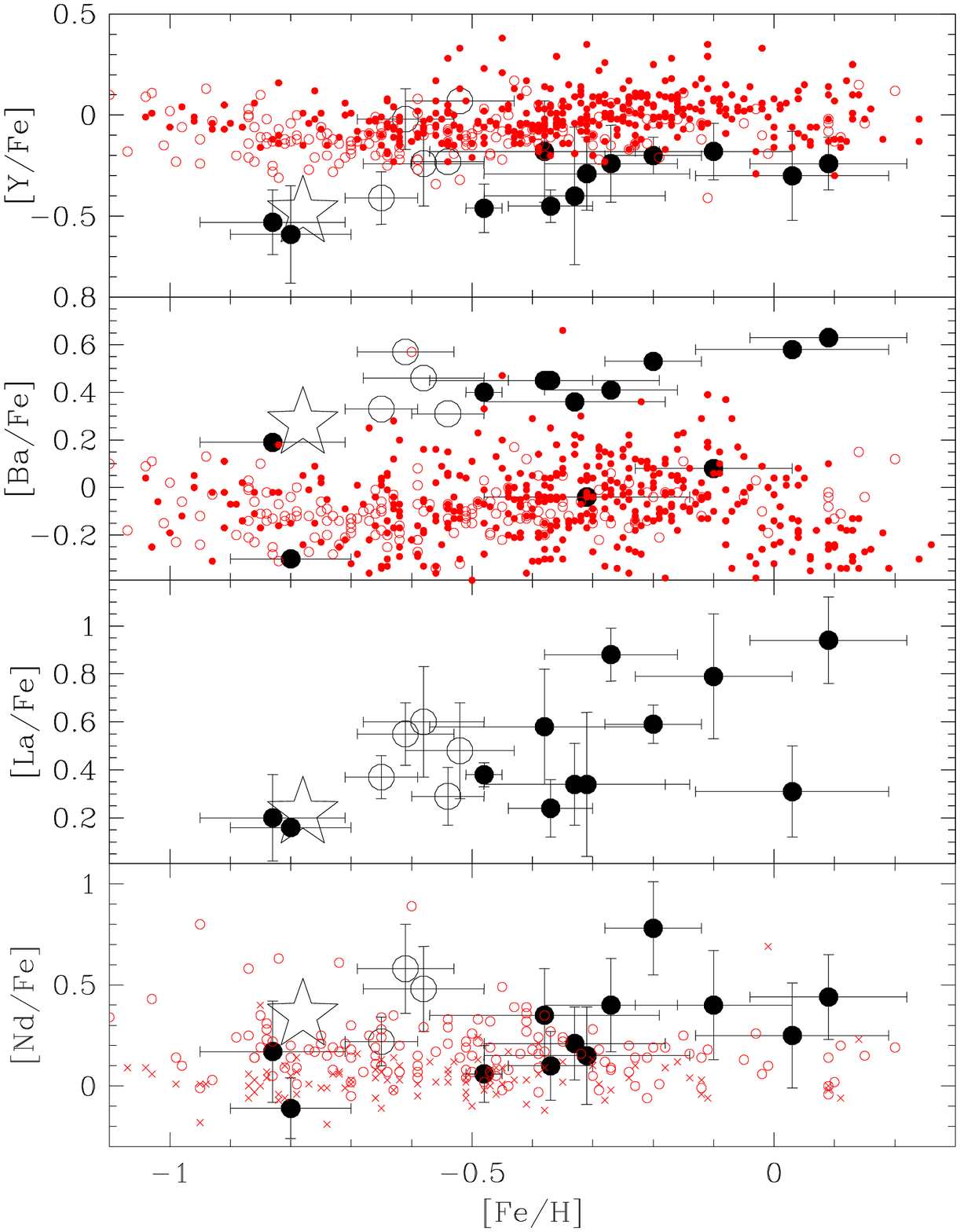}
\caption{[Y/Fe], [Ba/Fe], [La/Fe] and [Nd/Fe] plotted against [Fe/H].  Same symbols as in fig. \ref{NaAlScV} \label{YBaLaNd}}
\end{figure}

In figures \ref{alphasufe_tutti} through \ref{YBaLaNd} \object{Ter 7} stars are always represented by large open circles. As can be seen, \object{Ter 7} chemical composition appears to closely match the one observed in the \object{Sgr dSph} main body at corresponding metallicity. Nevertheless some differences can be seen: ratios against iron are slightly above the \object{Sgr dSph} values for V, Co, Ni, Y and Nd. This can be probably due to the lower reliability of the analysis in such low gravity -- low temperature atmospheres.

\subsection{Error budget}
\label{errors}

In table \ref{table_err} we report the variations in abundances, [X/Fe] and [Fe/H] for star \# \object{656}, {\em as due to the variation in atmospheric parameters between Paper I / II and this work}. This allows to estimate the impact of systematic uncertainties in the model parameters. Star \# \object{656} was the one with the largest temperature variation in the sample ($\Delta$T$_{eff}=-396$K, $\Delta \log{g}=-0.4$, $\Delta \xi = - 0.15$ \kms, new - old). Despite such a large variation in the parameters, [X/Fe] values show remarkably low changes.

\begin{table}[t]
\begin{center}
\caption{Variations in the derived abundances [X/Fe] and [Fe/H] for star \# \object{656} due to the change in atmospheric parameters between this work and Paper I, II. Absolute abundances from Paper II and this work are also listed to ease the comparison. Na abundances are listed here {\em without} the NLTE correction used elsewhere.  \label{table_err}}
{
\begin{tabular}{lllllr}
\hline
{\bf ion    }  & {\bf A(X)}         &{\bf A(X) }     & &{\boldmath $\Delta$}{\bf A(X)} & {\boldmath $\Delta$}{\bf [X/Fe]} \\
               & {\bf Paper I / II} &{\bf this work} & &{\bf new - old}                & {\bf new - old}                  \\
\hline
\\
\ion{O}{i}              & 8.50 & 8.15 & & -0.35 & -0.15 \\
\ion{Na}{i}             & 5.80 & 5.62 & & -0.18 &  0.02 \\
\ion{Mg}{i}             & 7.10 & 6.99 & & -0.11 &  0.09 \\
\ion{Al}{i}             & --   & --   & & --    & --    \\
\ion{Si}{i}             & 7.31 & 7.25 & & -0.06 &  0.14 \\
\ion{Ca}{i}             & 6.09 & 5.96 & & -0.13 &  0.07 \\
\ion{Sc}{ii}            & 2.55 & 2.34 & & -0.21 &  0.06 \\
\ion{Ti}{i}             & 4.78 & 4.44 & & -0.34 & -0.14 \\
\ion{V}{i}              & 3.80 & 3.38 & & -0.42 & -0.22 \\
\ion{Cr}{ii}            & 5.55 & 5.45 & & -0.10 &  0.17 \\
\ion{Mn}{i}             & 5.08 & 4.75 & & -0.33 & -0.13 \\
\ion{Fe}{i}             & 7.33 & 7.13 & & -0.20 & --    \\
\ion{Fe}{ii}            & 7.39 & 7.12 & & -0.27 & -0.07 \\
\ion{Co}{i}             & 4.59 & 4.07 & & -0.52 & -0.32 \\
\ion{Ni}{i}             & 5.77 & 5.51 & & -0.26 & -0.06 \\
\ion{Cu}{i}             & 3.44 & 3.05 & & -0.39 & -0.19 \\
\ion{Zn}{i}             & 3.94 & 3.92 & & -0.02 &  0.17 \\
\ion{Y}{ii}             & 1.93 & 1.68 & & -0.25 &  0.02 \\
\ion{Ba}{ii}            & 2.28 & 2.20 & & -0.08 &  0.18 \\
\ion{La}{ii}            & 1.63 & 1.33 & & -0.30 & -0.03 \\
\ion{Ce}{ii}            & 1.65 & 1.29 & & -0.36 & -0.09 \\
\ion{Nd}{ii}            & 1.86 & 1.47 & & -0.39 & -0.12 \\
\\
\hline
\end{tabular}                
}
\end{center}                 
\end{table}                  

\subsection{Comparison with other results}

The only  high resolution abundance analysis in the \object{Sgr dSph} independent of
those of our group, is the one presented in \citet{mcwilliam03,mcwilliam03err,mcwilliam05apjl,mcwilliam05ASP}. The authors analyze a sample of 14 \object{Sgr dSph} main body giants, with higher luminosity and lower temperatures than the ones presented din this work. The bulk  of their sample lies in the same metallicity range we explore here, while three stars lie below [Fe/H]=-1. Despite the difference in stellar parameters and in the details of the performed analysis, the results are largely coincident: Al, Na and Cu undersolar ratios are detected, aside with a positive [La/Fe], with increasing trend with the metallicity. The authors derive slightly higher [$\alpha$/Fe] ratios than we do, but not by a significant amount. The undersolar [Mn/Fe] ratio detected in \citet{mcwilliam03,mcwilliam03err} is also confirmed by our results, but the lack in our sample of stars below [Fe/H]=-1 does not allow us to confirm also the presence of an increasing trend of [Mn/Fe] with metallicity. In \citet{mcwilliam05apjl} the bulk of the analyzed stars show [Cu/Fe]$\sim$ -0.5, coincident with what we find. This is interesting since a different \ion{Cu}{i} feature is sampled with respect to the one we use. However four stars show significantly higher Cu content ([Cu/Fe]$\sim$0.3), something we do not detect.

The composition of three of our 5 \object{Ter 7} giants (\# \object{1282}, \#\object{1665} and \# \object{1708}) has been studied by \citet{tarantella04}. We already compared our results with the ones of \citet{tarantella04} in Paper III regarding atmosphere parameters, Iron and $\alpha$ element abundances. Regarding the other elements, while many the overall trends match (undersolar Na and Al, oversolar La...), some are at odd with our results (Sc, V, Mn, Co, Cu among others). As already stated in Paper III, the atmosphere parameters are fully consistent among us and \citet{tarantella04}, thus these discrepancies should be due to differences in the line data employed. Nevertheless, the fact that \citet{tarantella04} perform a differential analysis makes the comparison quite difficult.

\subsection{The long lost \object{Pal 12}}

In figures \ref{alphasufe_tutti} through \ref{YBaLaNd}, the average value of \object{Palomar 12} stars abundances from \citet{cohen04} are shown  as large open star symbols. By comparing these results against the limited sample of \object{Sgr dSph} abundances available at that time, \citet{cohen04} already deduced that the strong similarities between the two objects pointed towards a formation of \object{Palomar 12} within the \object{Sgr dSph} system. With the present work, the resemblance between \object{Pal 12} and \object{Sgr dSph} appears even more clearly: \object{Pal 12} displays precisely the same ``chemical signature'' encountered within \object{Sgr dSph} even in its most pronounced anomalies, such as the strong Na, Al, Co, Ni Cu and Zn underabundances. This leads to two main results: first, {\em the origin of \object{Pal 12} within the \object{Sgr dSph} can be considered as finally established}. There is no known trace of chemically similar populations inside the MW. It can be hypothized that another (now destroyed) dwarf galaxy may have dropped \object{Pal 12} within the Halo, but such object should have been  a sort of ``twin'' of \object{Sgr dSph} from an evolutive point of view. The proximity of \object{Pal 12} to the \object{Sgr dSph} stream in parameter space is also well known. The second, possibly more intriguing result, is that {\em looking for a Sgr-like ``signature'' in the chemistry of a stellar population is a very effective tool to search for \object{Sgr dSph} debris within the MW}, at least at not too low metallicity. Many of the \object{Sgr dSph} chemical oddities appear at any metallicity within the explored range, but it cannot be safely inferred what would happen of them at, say, [Fe/H]=-2. The $\alpha$ elements trend seems to resemble the one in MW populations below [Fe/H]=-1 (see fig. \ref{alphasufe_tutti}), while [Cu/Fe] is known to drop in the MW below [Fe/H]=-1. Further analysis of low-metallicity \object{Sgr dSph} stars is needed to clarify this.

\balance

\section{Discussion and Conclusions}

From the observational point of view, our results can be summarized as follows:

\begin{enumerate}

\item The chemical composition of \object{Sgr dSph} main population is significantly at odds with respect to the one observed within the MW. Many elements (Na, Al, Sc, $\alpha$\ elements, Co, Ni, Cu, Zn...) show significant underabundances when compared to MW stars of similar iron content. Such chemical oddities are exactly replicated within the associated globular cluster \object{Terzan 7}.

\item Even more interesting, precisely the same ``chemical signature'' is displayed by \object{Palomar 12}. This, as already noted by \citet{cohen04} essentially proves that this globular cluster originated within the \object{Sgr dSph} system to be subsequently stripped by the MW.

\end{enumerate}

From the phenomenological point of view, this has one main consequence: chemical composition should allow to distinguish also
{\em other} stellar populations which were stripped from \object{Sgr dSph} and added to the MW, at least for metallicities above [Fe/H]=-1. For lower metallicities, the survival of many of the chemical ``markers'' should be confirmed, but may be inferred in many cases by the absence of metallicity trends within them (e. g. Na, Ni, Zn). 

It has nevertheless to be taken into account the absence, in the known MW populations, of stars of Sgr-like characteristics. Recently \citet{chou06} claimed the detection of metallicity gradients within the \object{Sgr dSph} stream, in the sense of higher metallicities in the populations stripped more recently from the galaxy. From this finding, we can infer that a significant chemical evolution took place in the galaxy since the most distant stream stars were stripped, or (more likely) that the outermost parts of \object{Sgr dSph} (the first to be stripped) were significantly more metal poor than the nucleus. A combination of both scenarios is also possible. In all these cases, the ``chemical signature'' we detect in the main body of \object{Sgr dSph} may well be much weaker, or undetectable in the Halo at metallicities below [Fe/H]=-1. Conversely, \object{Palomar 12} stands as an unmistakable example of how evident this signature can be in more chemically evolved stars.

It appears a much harder task to decrypt the meaning of these abundances in terms of the chemical history of \object{Sgr dSph}, and we will not attempt to go deep in this regard in the present paper. \object{Sgr dSph} appears to have experienced a very long star formation and chemical evolution history. The lowest metallicity observed within the galaxy is about [Fe/H]=-3 \citep{zaggia04}, while the associated globulars (\object{M 54}, \object{Ter 7}, \object{Ter 8}, \object{Arp 2}, \object{Pal 12}) are dispersed between [Fe/H]=-2 and [Fe/H]=-0.6. At least the last few GYr of the galaxy evolution have taken place within the strong MW tidal field, which should have heavily influenced both its star formation and its yields retention capability. This may help explaining both the high mean metallicity of \object{Sgr dSph} and its scarce gas content in spite of having formed stars until a relatively recent past (likely a couple of GYr, see Paper II). At the same time, this may call for past (pre-interaction) \object{Sgr dSph} as being a relatively large, star forming, gas-rich object with a nucleated structure \citep{monaco05cusp} and a rich set of globular clusters.

The detailed abundance ratios we present need to be interpreted in the framework of a detailed chemical evolution model taking into account galactic winds \citep[such as in][]{lanfranchi03,lanfranchi04,lanfranchi05,lanfranchi06}. Galactic winds appear to play a key role in the evolution of small systems such as dSph, where they develop thanks to the relatively weak gravitational field these small galaxies create. Nevertheless, they still are more inferred than modeled, due to the great difficulty of the hydrodinamical calculations that would be needed. In the particular case of \object{Sgr dSph}, the influence of the MW tidal field likely favored the wind formation, an effect that has been most likely {\em time dependent}, since \object{Sgr dSph} orbit should have degraded with time. 

Aside from the quoted issues, we want to stress how the presented abundances should not be taken as plainly representative of \object{Sgr dSph} ``as a whole''. This not only comes from the small range in metallicity covered, but also from the small angular area from which they have been collected. \citet{chou06} findings point towards a scenario in which pre-interaction \object{Sgr dSph} may have showed strong metallicity gradients, or maybe even an incomplete chemical mixing. Especially at low metallicities, chemical enrichment may have been locally influenced by a small number of SN II, due to the low total mass of the galaxy. Larger sampling across the galaxy body and the streams are needed, as well as analyses of the more metal poor components, to be able to trace the full story of this fascinating neighbor.

\begin{table*}[p]
\caption{Absolute abundances, assumed solar abundances, [Fe/H], [X/Fe] values and associated errors for \object{Sgr dSph} main body stars \# \object{432} \# to \object{716}. The upper panel hosts assumed solar abundances and absolute measured abundances for the sample stars. The published error is simply the RMS of the measured lines, thus is absent for species where a single line has been used. The lower panel lists [X/\ion{Fe}{i}] ratios for the neutral species, [X/\ion{Fe}{ii}] ratios for the ionized ones. [O/\ion{Fe}{ii}] is listed instead of [O/\ion{Fe}{i}] due to the strong gravity sensitivity of {[\ion{O}{i}]} 630.03 nm line. Iron values are [Fe/H] under \ion{Fe}{i} and [\ion{Fe}{ii}/\ion{Fe}{i}] under \ion{Fe}{ii}.}\label{abs_abu_sgr_1}
\centering
{\scriptsize
\begin{tabular}{llllllllllllll}
\hline
 & {\bf Sun} & {\bf \object{432} } &                     & {\bf \object{628} } &                     & {\bf \object{635}} &                     & {\bf \object{656}} &                     & {\bf \object{709}} &                     & {\bf \object{716}} &                       \\
 & {\bf A(X)}& {\bf A(X) }         & {\boldmath $\sigma$}& {\bf A(X) }         & {\boldmath $\sigma$}& {\bf A(X) }        & {\boldmath $\sigma$}& {\bf A(X) }        & {\boldmath $\sigma$}& {\bf A(X) }        & {\boldmath $\sigma$}& {\bf A(X) }        &  {\boldmath $\sigma$} \\
\hline
\\
\ion{O}{i}      & 8.72 & 8.03 & --   & 8.35 & --   & 8.33 & --   & 8.15 & --   & 8.50 & --   & 8.54 &  --   \\
\ion{Na}{i}     & 6.33 & --   & --   & 5.81 & 0.08 & 5.40 & 0.03 & 5.74 & 0.08 & 6.03 & 0.13 & 5.83 &  0.16 \\
\ion{Mg}{i}     & 7.58 & 6.64 & 0.01 & 7.16 & 0.12 & 7.10 & 0.13 & 6.99 & 0.04 & 7.21 & 0.12 & 7.13 &  0.02 \\
\ion{Al}{i}     & 6.47 & --   & --   & --   & --   & --   & --   & --   & --   & --   & --   & 5.61 &  --   \\
\ion{Si}{i}     & 7.55 & 6.66 & 0.17 & 7.30 & 0.10 & 7.12 & 0.11 & 7.25 & 0.08 & 7.37 & 0.14 & 7.39 &  0.15 \\
\hline                                                                                                                       
\end{tabular}                                                                                       
}
\end{table*}                                                                                

\begin{table*}[p]
\caption{Same as in table \ref{abs_abu_sgr_1} for stars \# \object{717} to \object{927}. \label{abs_abu_sgr_2}       }
\centering
{\scriptsize
\begin{tabular}{llllllllllllll} 
\hline
	& {\bf Sun} & {\bf \object{717} } & 	& {\bf \object{772} } & 	& {\bf \object{867}} & 	& {\bf \object{879}} & 	& {\bf \object{894}} & 	& {\bf \object{927}} & 	\\
	& {\bf A(X)}& {\bf A(X) } & {\boldmath $\sigma$}& {\bf A(X) } & {\boldmath $\sigma$}& {\bf A(X) } & {\boldmath $\sigma$}& {\bf A(X) } & {\boldmath $\sigma$}& {\bf A(X) } & {\boldmath $\sigma$}& {\bf A(X) } & {\boldmath $\sigma$} \\
\hline
\\
\ion{O}{i}      & 8.72 & 8.49 & --   & 8.07 & --   & 8.01 & --   & --   & --   & 8.33 & --   & 8.45 & --    \\
\ion{Na}{i}     & 6.33 & 5.87 & 0.16 & 5.52 & 0.15 & 5.36 & 0.07 & 5.64 & 0.08 & 5.82 & 0.10 & 6.00 & 0.14  \\
\ion{Mg}{i}     & 7.58 & 7.33 & 0.05 & 6.85 & 0.13 & 6.76 & 0.11 & 6.96 & 0.13 & 6.98 & 0.16 & 7.08 & 0.09  \\
\ion{Al}{i}     & 6.47 & 5.66 & --   & --   & --   & --   & --   & 5.76 & --   & --   & --   & 5.70 & --    \\
\ion{Si}{i}     & 7.55 & 7.51 & 0.14 & 7.18 & 0.12 & 6.96 & 0.15 & 6.98 & 0.03 & 7.41 & 0.07 & 7.40 & 0.08  \\
\hline
\end{tabular}
}
\end{table*}

\begin{table*}[p]
\caption{Same as in table \ref{abs_abu_sgr_1} but now for \object{Terzan 7} stars. \label{abs_abu_ter}}
\centering
{\scriptsize
\begin{tabular}{llllllllllll}
\hline
	& {\bf Sun} &  {\bf \object{1272}} & & {\bf \object{1282}} & 	& {\bf \object{1515}} & 	& {\bf \object{1665}} & 	& {\bf \object{1708}} \\
	& {\bf A(X)} &  {\bf A(X) } & {\boldmath $\sigma$}& {\bf A(X) } & {\boldmath $\sigma$}& {\bf A(X) } & {\boldmath $\sigma$}& {\bf A(X) } & {\boldmath $\sigma$}& {\bf A(X) } & {\boldmath $\sigma$} \\	
\hline
\\
 \ion{O}{i}     & 8.72 & 8.17 & --   & 8.36 & --   & --   & --   & 8.43 & --   & 8.34 & --   \\
 \ion{Na}{i}    & 6.33 & 5.49 & 0.02 & 5.48 & 0.06 & 5.42 & 0.04 & 5.49 & 0.07 & 5.32 & 0.04 \\
 \ion{Mg}{i}    & 7.58 & 6.88 & 0.13 & 6.88 & 0.05 & 6.99 & 0.09 & 6.93 & 0.12 & 6.84 & 0.08 \\
 \ion{Al}{i}    & 6.47 & 5.55 & 0.11 & 5.47 & 0.01 & 5.66 & --   & 5.43 & 0.02 & 5.43 & --   \\
 \ion{Si}{i}    & 7.55 & 6.91 & 0.16 & 7.06 & 0.11 & 7.14 & 0.05 & 7.00 & 0.12 & 6.98 & 0.20 \\
\hline
\end{tabular}
}	
\end{table*}

\begin{acknowledgements}
LS, PB and SZ acknowledge support from the MIUR/PRIN 2004025729\_002;
LS and PB are supported also by  EU contract MEXT-CT-2004-014265 (CIFIST). This work has made use of
NASA's Astrophysics Data System Bibliographical Services and of NIST online database of atomic data.
\end{acknowledgements}

\Online
\appendix

{\scriptsize
\begin{longtable}{rrrlrrrrrrrrrrrr}
\caption{Employed lines, $\log gf$ values, $\log gf$ sources, observed EW and derived abundances for \object{Sgr dSph} main body stars \# \object{432} to \# \object{716}. Codes for the source of the $\log gf$ can be found in the bibliography, NIST code referst to $\log gf$ derived from the NIST online database. For lines that have been synthesized, ``syn'' substitutes the EW; for lines which have been synthesized by using HFS, log gf is not provided, substituted by ``HFS'', and wavelength is truncated to 0.1 nm. Pure LTE abundances, without NLTE corrections, are published here for Na lines.}\\
\hline\\
Ion        & $\lambda$ & $\log gf$  & source of    & EW      & A(X)       & EW       & A(X)       & EW      & A(X)       & EW      & A(X)       & EW     & A(X)       & EW     & A(X)      \\
           & (nm)      &         & $\log gf$       & (pm)    &            & (pm)     &            & (pm)    &            & (pm)    &            & (pm)   &            & (pm)   &           \\
           &           &         &              & \object{432}     &            & \object{628}      &            & \object{635}     &            & \object{656}     &            & \object{709}    &            & \object{716}    &           \\
\\
\hline
\\
\ion{O }{i} & 630.0304 & -9.717  & STZ          &   syn   &  8.03      &   syn    & 8.35       &   syn    & 8.33      &   syn   &  8.15      &  syn   &  8.50      &  syn   &  8.54     \\
\hline
\end{longtable}

\begin{longtable}{rrrlrrrrrrrrrrrr}
\caption{As in previous table, for \object{Sgr dSph} main body stars \# \object{717} to \# \object{927}.}\\
\hline\\
Ion        & $\lambda$ & $\log gf$  & source of    & EW      & A(X)       & EW       & A(X)       & EW      & A(X)       & EW      & A(X)       & EW     & A(X)       & EW     & A(X)      \\
           & (nm)      &         & $\log gf$       & (pm)    &            & (pm)     &            & (pm)    &            & (pm)    &            & (pm)   &            & (pm)   &           \\
           &           &         &                & \object{717} &        & \object{772} &        & \object{867} &       & \object{879} &       & \object{894} &      & \object{927}    &           \\
\\
\hline
\\
\ion{O }{i} & 630.0304 & -9.717  & STZ          &  syn    &  8.49      & syn      &   8.07     &  syn    &   8.01     &  --     &   --       & syn    &   8.33     &  syn   &   8.45    \\
\hline 
\end{longtable}

\begin{longtable}{rrrlrrrrrrrrrr}
\caption{Same as in previous tables, now for the \object{Ter 7} stars.}
\\
\hline\\
Ion  & $\lambda$ & $\log gf$  & source of   & EW      & A(X)       & EW       & A(X)       & EW      & A(X)       & EW      & A(X)       & EW      & A(X)      \\
     & (nm)      &         & $\log gf$      & (pm)&            & (pm) &            & (pm)&            & (pm)&             &(pm)&           \\
     &           &         & (see notes) & \object{1272}    &            & \object{1282}     &            & \object{1515}    &            & \object{1665}    &             & \object{1708}   &           \\
\hline\\
\ion{O}{i}   & 630.0304 & -9.717 & STZ    & syn   & 8.17  & syn   & 8.22  & syn   & --    & syn   & 8.20  & syn & 8.24 \\
\\
\hline

\end{longtable}
}

\end{document}